# Causality, light speed constancy and local action principle.

Enrique Ordaz Romay[1]

*Facultad de Ciencias Físicas, Universidad Complutense de Madrid*

## Abstract

In 1905 A. Einstein [1], from the experiments of Michelson and Morley in 1887 [2], enunciates the light speed constancy principle in the inertial frames of reference. However, this principle was pointed by the equations of the electromagnetism of J. C. Maxwell in 1865 [3]. The prediction of the Maxwell equations is due to in the most basic principle of the science, the principle of causality, is the reason of the light speed constancy. This article shown how, only with the causality principle, we can deduce the properties of the light. At the end of the article we used the causality relativistic definition to approach the local action principle, as well as to solve the paradox that implies the EPR theorem.

[1] eorgazro@cofis.es

# Introduction.

The restricted relativity is based on the traditional principles of the classic physics and the light speed constancy principle in the inertial frames of reference. Such traditional principles of the classic physics are, basically [4§1]:

1. The tridimensionality of the longitudinal space.

2. The existence of the time dimension (together to its absolute character that gives it the classic physics and to the relative that gives it the relativity).

3. The existence of continuous physical magnitudes.

4. The action principle (Principle of Least Action in other authors).

While the first three principles are based on the direct observation, the action principle is based on the experience, this is, in the accumulation of observations and in the deduction of common principles for such observations [5§2].

The first principle that the experience teaches us is the principle of causality [6], according to which "all event are determined by the history of previous events. To these previous events we called causes and to the event that they determine we called effect." [7].

This article will show how the light speed constancy principle in the inertial frames of reference can be deduced from the principle of causality applied to the homogeneous spaces. The inverse process, that is, beginning from the light speed constancy and to arrive to the causality approaches, is frequent in the texts [8]. However, these developments suppose that we arrive to a more evident event: the causality, starting from a more "laborious" principle: the light speed constancy.

If beginning from the causality we can concluding the light speed constancy, this would imply that the relativity is an innate characteristic of the causality and not necessarily a posteriori consequence of an experimentation. Although, we do not seek, at all, to subtract the importance to the experimental physics as inspiration and confirmation or refutation of the hypotheses of the theoretical physics.

To show this dependence between both principles (light speed constancy regarding the causality) we will begin defining the physical space, later the homogeneous space and deducing its properties starting from its definition. Lastly we will apply the principle of causality in the homogeneous space and we will obtain the light speed constancy principle in the inertial frames of reference.

At the end of this article we will generalize the principle of causality to make it compatible with a relativistic idea (restricted or general), according to this, the time coordinate is in equality with the longitudinal coordinateses. This will help us to understand the local action principle and to solve the paradox of the EPR theorem.

## The physical space.

The physical space is characterized by:

1. The tridimensionality of the longitudinal space.

2. The existence of the time.

In a general vision we can observe that, the absolute character that concede the classic physics to the time, is a particular case of a relative time [8]. We can start from the supposition of the time is relative and if in the coordinate transformations and times, from some reference frames to other, the time always has the same value in all the reference frames, we will conclude that the time is absolute. On the contrary, if we find that the time also change in the transformation from a reference frame to another, we will conclude that the time is, indeed, relative.

According to this reasoning, the two previous observations can become in:

1'. The physical space is formed by four dimensions: three space dimensions and one time dimension.

According to this first observation (1') the numeric set on which is defined the physical space will be the Cartesian product of four real number sets or a subset of this. Mathematically, this is symbolized by:

$$\Omega \subset \Re^4 = \{X / X = (x^0, x^1, x^2, x^3), \forall x^i \in \Re\}$$

To $\Omega$ it is called physical system or system simply.

Just as the mathematical analysis teach, in this numeric set we can defined a distance function between two points [9]. It is said that a function $s$ is a distance between two points $X, Y \in \Omega$, if it is an application $s : \Omega \otimes \Omega \to \Re$, which verifies the following properties:

I. It is only zero the distance from any point with itself: $s(X,Y) = 0 \Leftrightarrow X = Y$.
II. The distance is symmetrical: $s(X,Y) = s(Y,X)$.
III. It verifies the triangular inequality: $s(X,Y) + s(Y,Z) \geq s(X,Z)$.

Let the covectors set $\{e_i\} \subset \Omega$ be a ortonormal basis from $\Omega$. Then, any point $X$ (which defines to the covector $\overline{OX}$) it will be represented by $X = x^i e_i$ [10]. For the properties of the distance, the expression $(ds)^2$ is symmetrical, bilineal and defined positive. In this way, we can write that the distance between two infinitesimally near point $X + dX$ and $X + dX'$ adopts the expression:

$$(ds)^2 = g(X + dX, X + dX') = g(dX, dX') = dx^i g(e_i, e_j) dx'^j$$

being the function $g$ a symmetrical and bilineal function. This function is represented by a matrix $G = (g_{ij})$ called metric matrix which components have the form $g_{ij} = g(e_i, e_j)$. From this notation the distance expression becomes:

$$(ds)^2 = dx^i g_{ij} dx^j = dX^+ \cdot G \cdot dX$$

The superindex $^+$ represents the transpose and complex conjugated transformation (as until the moment we handle real numbers, the complex conjugation is not significant).

As $G$ is symmetrical and its determinant is different to zero, the equation of diagonalization $\|G - \lambda I\| = 0$ is verify for no degenerates values of $\lambda$. That is to say, for the four vales of the diagonal.

On the other hand if $\{v_i\}$ are the vectors of the ortonormal basis of diagonalization, then, according to the diagonalization theorem [9], the matrix $T$, formed by the columns of the vectors of the basis, verifies the equation: $G = T^{-1} G_d T$.

As well the normal metric as the diagonal are symmetrical, the expression: $G^+ = (T^{-1} G_d T)^+ = T^+ G_d (T^{-1})^+$ should be similar to $G = T^{-1} G_d T$ and this is possible only if $T^{-1} = T^+$. Therefore $T$ is an unitary matrix [11].

In summary, it exist $T : \Re^4 \to \Re^4$ a biyective transformation $X' = T(X)$, provided that $G = T^+ G_d T$ with $G_d$ diagonal. That is to say: $G_d = \begin{pmatrix} \lambda_0 & 0 & 0 & 0 \\ 0 & \lambda_1 & 0 & 0 \\ 0 & 0 & \lambda_2 & 0 \\ 0 & 0 & 0 & \lambda_3 \end{pmatrix}$. $G_d$ is called diagonalized metric.

## The homogeneous space.

It is said that a physical system is homogeneous when "the value of any physical magnitude is independent of the 4point of the system where we measure it"[2]. Concretely, the

---

[2] Any dictionary defines the homogeneous space as a space in which the physical magnitudes are identical in any point of the 4space. That is to say, their values are constants and independents of the point in which we are measured. That is the definition that we use there. If we want measure a magnitude we used a standard of this magnitude, to compare it with the magnitude that we want to measure and to extract the proportion that exists between the standard and the object of the measure. However, when we try to apply this process to the

metric defined in it is not function of the coordinateses of the point where it is applied. That is to say:

$$\frac{\partial G}{\partial x^i} = \partial_i G = 0$$

If $\partial_i G = 0$ then $G = constant$. Therefore, the components and the diagonalization values of $G$ are constant. That is to say, when the physical system is homogeneous, the components of the metric matrix and of the diagonal metric matrix are constant in all the system.

For tradition we will denote, in Cartesian coordinates, to $X^0$ as the time coordinate and to $X^1, X^2, X^3$ as the longitudinal three coordinate of each point of the 4space [12].

For verify the implicit isotropic in the homogeneous systems it is necessary that any reference frame, in which $G$ remains constant, produce the same results. Like this, if we exchange the space coordinates, for example, the axis $X^1$ with the $X^2$, the $X^2$ with the $X^3$ and the $X^3$ with the $X^1$, will stay unaffected the homogeneity of the system. Therefore, the space coordinates in a homogeneous system are interchangeable and the homogeneity of the system does not alter.

Following this reasoning we will look for the form of the diagonal metric matrix. For this, let us to analyse what is the meaning of the space coordinates can be interchangeable.

---

homogeneous space we find a problem. As by definition, in the homogeneous space, the physical magnitudes do not depend of the coordinateses, then the introduction of a magnitude standard in the system breaks the homogeneity and perturbs our system. In this way, it is not easy to understand the characteristics of the homogeneous space, because its own concept annuls the existence of differentiating characteristic. In the homogeneous space the magnitudes are distributed in so uniform form that it is impossible their detection. In this way, in the homogeneous space, all the magnitudes possess the same value in all de points, although we cannot know which is this value or if, in fact, this value is zero. Since, in the homogeneous space differentiating characters do not exist among a point and another, all the regions of the system have to be same, independently of the coordinated transformation that we use to pass from a region to another one.

Let us take the two first longitudinal coordinateses, for example $X^1, X^2$. As they are interchangeable then, it exists $f: X^1 \to X^2$ a lineal application such that

$$x^2 = f(x^1) \Rightarrow dx^2 = \frac{df(x^1)}{dx^1} dx^1$$

and reciprocally $dx^1 = \frac{df^{-1}(x^2)}{dx^2} dx^2$. Substituting in the expression of $G_d$, the distance takes the form:

$$ds^2 = \lambda_1 \left( \frac{df^{-1}(x^2)}{dx^2} \right)^2 (dx^2)^2 + \lambda_2 \left( \frac{df(x^1)}{dx^1} \right)^2 (dx^1)^2 + ...$$

So that the metric keeps constant although the coordinateses are interchangeable, it should be verify that $\lambda_1 = \lambda_2 \left( \frac{df(x^1)}{dx^1} \right)^2$. This implies that $\frac{df(x^1)}{dx^1} = cte$, that is to say, the $f$ function is a lineal function. If we take the linearity constant similar to one, $f$ maintain the form[3] $f(x) = x$ and in consequence it is that $\lambda_1 = \lambda_2$. The function $f(x) = x$ used in the application $f: X^1 \to X^2$ is in fact the transformation that identifies, to each point $X^1$, with one point $X^2$ whose numeric value coincides.

In consequence, the exchange of the first two longitudinal coordinateses it drives to that $\lambda_1 = \lambda_2$.

Applying this same reasoning with the third coordinate we concludes that, in a homogeneous system, the appropriate election of units completes $\lambda_1 = \lambda_2 = \lambda_3$. To this value we will simply denote it for the letter $\lambda$ so that the metric adopt the form:

$$(ds)^2 = \lambda_0 (dx^0)^2 + \lambda \left[ (dx^1)^2 + (dx^2)^2 + (dx^3)^2 \right] \qquad (1)$$

---

[3] If we took the linearity constant different to one this only would suppose a difference in the units system in which we are measured both coordinateses.

# The principle of causality (1).

The principle of causality tells us that all event is determined by the history of the previous events. These previous events are called causes and the event which is determine is called effect [7].

We will call event to a happening that fact in a point of the 4space and we will represent it for the coordinateses of this 4point. According to the principle of causality, an event, in a region of the space, will be integrated by the punctual events that it contains the region.

The observation of the principle of causality is so essential because it is verify independently of the referencial frame that we use. That is to say, if an event is cause of another one in a frame, it will be in any other frame, too.

Applying this observation in the homogeneous[4] space we can find the relationship that exists among $\lambda_0$ and $\lambda$. For this, it is necessary to make some reflections and to introduce some concepts.

So that an event is cause of another, it is necessary that, when it is produce the second event, this "has had news" of the first one. That is to say, the information referred to the first event should have arrived to the 4point where the second event will produce, before this second event produce it.

So that the principle of causality was verified in any reference frame, this fact should be independent of the coordinated system that we use, that is to say, the interval or 4dimensional distance between two events are constant and independent of the reference frame.

---

[4] Although in a homogeneous system cannot be given events because, the differentiating singularity of these, would break the homogeneity of the system. We are not interested in the events in their self, but in the causal relationship among the events. For this, we can be strict and to consider the events outside of our homogeneous space.

Now, we have a clear relationship between the space dimensions and the time dimension. To analyze the physical consequences of the principle of causality is necessary to define the concept of 3-distance or longitudinal distance, 3-speed or three-dimensional speed and 3-speed of the information.

- Longitudinal distance three-dimensional, in Cartesian coordinates and in a homogeneous system is defined as $dl^2 = (dx^1)^2 + (dx^2)^2 + (dx^3)^2$.
- The three-dimensional speed is defined as $\dfrac{dl}{dx^0}$ when $l$ is the cover distance and $x^0$ is the time that takes a long in cover this distance.
- The three-dimensional speed of propagation of a signal (of the "information") is defined as $\dfrac{dl_{inf}}{dx^0_{inf}} = c$ when $l_{inf}$ is the distance cover by the signal (information) and $x^0_{inf}$ is the time that takes in cover this distance.

Let us two events $A$ and $B$ such that, when the signal of the event $A$ arrives to the three-dimensional coordinateses of $B$ is when the event $B$ takes place. Under these conditions the longitudinal distance from $A$ to $B$ is similar to $l_{inf}$ and the time from $A$ to $B$ is $\Delta x^0_{inf}$. According to this,

$$c^2(x^0_A - x^0_B)^2 = (x^1_A - x^1_B)^2 + (x^2_A - x^2_B)^2 + (x^3_A - x^3_B)^2 \Rightarrow c^2(\Delta x^0_{inf})^2 = (\Delta l_{inf})^2 \qquad (2)$$

Calculating now, in the homogeneous system, the infinitesimal 4distancia from $A$ to $B$ (1) to which we called interval and substituting (2) in it, we find

$$(ds_{AB})^2 = (\lambda_0 + \lambda c^2)(dx^0_{inf})^2 \qquad (3)$$

As $ds^2$, $(dx^0)^2$ and $(dl)^2$ are lineal in the homogeneous systems and when $A$ tend to $B$ both the interval between $A$ and $B$ and the difference in the coordinateses between the both points tend to zero, from (3) we arrive to:

$$\lambda c^2 = -\lambda_0 \qquad (4)$$

From (4) we can conclude several important results:

1. As in the homogeneous space $\lambda$ and $\lambda_0$ are constant, the three-dimensional speed to which the signals (information) propagate is the same in all the frames in which the space is keep homogeneous. To these frames we called it inertial frames.

2. The second result tells us that the signs (plus/minus) of $\lambda$ and $\lambda_0$ are contrary. That is to say, if in our units system the magnitude of the signal (information) speed is the unit, then $-\lambda = \lambda_0$. As making $c$ equal to the unit does not determine the units system completely but only the relationship between the time unit and the space one, then we will even be able to find the units system in which $-\lambda = \lambda_0 = 1$.

3. Lastly, the interval between two events which three-dimensionally distance is the distance that cover the signal (information) from one to the events to the other one, is similar to zero. That is to say $s_{AB} = 0$

In this form, according to the point number 2, the metric of the homogeneous space has the form:

$$G(0) = \begin{pmatrix} 1 & 0 & 0 & 0 \\ 0 & -1 & 0 & 0 \\ 0 & 0 & -1 & 0 \\ 0 & 0 & 0 & -1 \end{pmatrix}$$

This metric matrix is called Minkowski metric [13] and the space invested with this metric is called Minkowski space.

However, the expression $-\lambda = \lambda_0$ that we have used pose a problem. The interval, between two any points in the 4space, does not behave like a distance in the all space (since it does not verify the property I of the distance), but only for points of a certain region.

Therefore, the physical space is not, in the mathematical sense, a metric space, but a pseudometric space.

Regarding the signal (information) speed for an homogeneous space we should observe that, if when a signal crosses the space it breaks the homogeneity, then it cannot complete the condition of constancy of the signal speed. This constancy is only valid in case the space remains homogeneous.

If to transmit a signal we send a particle that possesses a magnitude like mass or charge through a homogeneous space, the space loses its homogeneity and therefore the speed of the particle will not be a constant.

For transmit information and that the space does not lose the homogeneity, we need to use a signal that does not possess physical magnitudes, such as mass or charge. In this case, the signal will travel to an uniform speed for any inertial frame. To the signal that verify this characteristic it is called light. The energy that transport its particles without mass is its own undulate vibration and the spin is not more than a intrinsic property no susceptible of variation (as the photon does not have charge, the spin does not produce gyro magnetic field). When a signal of light crosses a homogeneous space, this continues being homogeneous. If we make discharge the energy that contained the signal of light on an object, evidently we break up the homogeneity with the object in question and the signal loses its uniform speed, however, during the time that was traveling the signal, as it did not break the homogeneity of the space, its speed could keep constant.

All that we have said until here is valid for any homogeneous space. A very important case of homogeneous space is the vacuum space. In the vacuum space it will be verify the constancy of the signals for the inertial frames, too. It is the call principle of constancy of the light speed in the vacuum [1].

## The Minkowski space

For a point *P*, the Minkowski space, just as we have defined it, is divided in three regions [12]:

1. $\{X\ /\ s(P, X) > 0\}$ temporal region: an object with mass can go from $P$ to any point of this region. $s\,(P, X)$ behaves like a distance for any value of $X$. To this interval we call it time interval.
2. $\{X\ /\ s\,(P, X\,) = 0\}$ light cone: only the light signal can go from $P$ to any point of this region. $s\,(P, X)$ does not behave like a distance. To this interval we call it null interval.
3. $\{X\ /\ s\,(P, X) < 0\}$ space region: an object with mass cannot accelerate to go from P to the $X$ point. However, $s\,(P, X)$ behaves like a distance, with negative values, for any value of $X$. To this interval we call it space interval.

According to this, the light cone is a authentic frontier. As the concept of interval is not an authentic 4dimensional distances, the mathematical analysis of a system that contains one part in each Minkowski region forces to make the analysis of the system from a previous 4point which contains all the system in the same region. This fact will be the key to solve the paradox of the EPR theorem.

The form in which transform the coordinateses between two inertial frames, in the Minkowski space, is obtained in the following form: Two inertial frames is characterized by, in both, the metric is a Minkowski metric. Be $S$ and $S\,'$ the two inertial frames. The metric of both are:

$$s^2(X_1, X_2) = X_1^+ \cdot G(0) \cdot X_2$$
$$s'^2(X'_1, X'_2) = X'^+_1 \cdot G(0) \cdot X'_2$$

Be $X\,' = T \cdot X$ the transformation that passes the coordinateses from $S$ to $S\,'$. Substituting in the second expression should obtain the first one. Operating is obtained that $T$ should verify

$$G(0) = T^+ \cdot G(0) \cdot T \qquad (5)$$

To the set of the matrix that verify this condition is called "Lorentz complete group" [14] and it is represented by $L$.

The condition (5) implies that $\det(T) = \pm 1$. It is denoted by $L_+$ and is called "Proper Lorentz group" to the matrixes $T$ that have determinant positive. These matrixes can break down as product of translations and rotations. An important case of translations is the Lorentz simplex matrix which form is:

$$L_W = \begin{pmatrix} a & \sqrt{a^2-1} & 0 & 0 \\ \sqrt{a^2-1} & a & 0 & 0 \\ 0 & 0 & 1 & 0 \\ 0 & 0 & 0 & 1 \end{pmatrix}$$

This matrixes belong to the proper Lorentz group and from them it is built all the Special Relativity and the transformations in the Maxwell electrodynamics.

## The principle of causality (2).

Until here we have based on two principles:

1'. The physical space is formed by four dimensions: three space dimensions and a time dimension.

2. The principle of causality tells us that all event comes determined by the history of previous events. These previous events are called causes and to the event which they determine is called effect.

and in a definition:

3. It is said that a physical system is homogeneous when the value of any physical magnitude is independent of the 4point of the system where we measure it.

The principle 1 ' is the unification of two previous principles which we have applied a generalization. This generalization consists on supposing that the time is a coordinate that is on equal footing with the three longitudinal coordinateses.

It is clear that, in the principle of causality, just as we observe it here, this generalization is not given for the time coordinate. The causality concede to the classification of the events in function of the time coordinate a bigger importance that to the classification in function of the space coordinates.

As we just to see, the causal relationship, can enunciate in the form: "Two events $A$ and $B$ can show causal relationship if its interval is bigger or the same as 0" [12]. Mathematically it would be expressed: $A$ and $B$ are causally connect if $(s_{AB})^2 > 0$. According to this, the four coordinateses are in equality with this definition of causal relation.

But in order that stay on equal footing, really, the coordinates space and time, to determine, if $A$ is cause of $B$ or vice versa, may be an other part than the consideration 4positional relative from an event regarding to another one.

It is well-know that, if we take the reference frame 4-centred in $A$ and we observe that $x^0(B) > 0$ we will say that $A$ can be cause of $B$. Reciprocally, if that $x^0(B) < 0$ we will say that $B$ can be cause of $A$.

To observe the relativity of this classification with regard to the time coordinate, we observe that, we can make these same classifications with regard to another coordinate. If, for example, $x^1(B) > 0$ can say that the event $B$ takes place on the right[5] from $A$, and vice versa, if $x^1(B) < 0$ will say that $B$ takes place on the left from A.

However, when the interval between $A$ and $B$ is temporal we will find one inertial frame in which $B$ is on the right from $A$ and another frame in which $B$ is on the left from $A$. On the other hand, when we say that $A$ can be cause of $B$, it does not exist one frame in which this fact can be changed. This should not miss us because it is the hypothesis that we have start: "the experience tells us that if $A$ is cause of $B$ it will be in any inertial frames." From there it arises that in the Minkowski metric the first term of the diagonal has the sign contrary to the rest of the elements of the diagonal.

It could seem, in a principle that, this fact radically changes the nature of the time in opposite of the space. However, the last analysis that we made, is not complete, (we have seen only events separate a time interval) and therefore it is soon to make such affirmation.

---

[5] Right = in positive way of $X^1$; Left = in negative way of $X^1$.

When the interval between *A* and *B* is space-like, the things change: as $(s_{AB})^2 < 0$, the events *A* and *B* do not have causal relationship. In some inertial frames *A* will be previous in the time to *B* and in other *B* will be previous in the time to *A*. However, the space coordinates behave in a different form: for example, if in an inertial frame *A* is on the right from B, it does not exist Lorentz simple transformation to another inertial frame in the one which *A* is on the left of *B*. So that we get this it is necessary to make a rotation of the reference frame so that during this rotation it leaves off to be inertial.

We can make the same reasoning with the time coordinate. A rotation of the frame that supposes a direction change in the time coordinate will make that, among events causally connected, the efects become causes and vice versa. This does not contradict our hypothesis about the causality, since the direction change in the time requires a no inertial transformation.

In summary:

1. If *A* and *B* is separated by a <u>time</u> interval:

    a. The relative <u>temporal</u> position between *A* and *B* can <u>NOT</u> change for inertial transformation. This is called causal relationship.
    b. The relative <u>space</u> position between *A* and *B* <u>YES</u> can change for inertial transformation.

2. If *A* and *B* is separated a <u>space-like</u> interval:

    a. The relative <u>temporal</u> position between *A* and *B* <u>YES</u> can change for inertial transformation. This would be called absence of causal relationship.
    b. The relative <u>space</u> position between *A* and *B* can <u>NOT</u> change for inertial transformation.

# The local action principle.

The local action, enunciated by A. Einstein and others in 1935 [15] says: "Two isolated systems, spatially separate, are independent."

This enunciated is used in the well-known theorem EPR [15] according to the one which: "or the wave function is a maximal representation of the system or it is valid the local action principle. But both cannot be valid at the same time."

Here we will show that the local action principle is not correct from its current enunciated, independently of the quantum considerations that we make, because the error is in its own one enunciated.

If two systems $A$ and $B$ are independent, it is necessary, not only that are causally unconnected, but something stronger: "it will not exist any systems that has causal connection so much with $A$ as with $B$." To this system, if exist, we will call it "common causal origin".

Be $S_A$ the action function that describes to the system $A$ and $S_B$ the one that describes to the system $B$. $A$ and $B$ are independent means that NO exists a generalized coordinate or function that it is common to both. That is to say, it does not exist $f$ such that, at the same time $S_A = S_A (f)$ and $S_B = S_B (f)$. If it existed this coordinate or function $f$, this would be the common causal origin. Therefore, the independence condition between both systems is the same as the condition of that both systems have not a common causal origin.

The local action principle talk about isolated and spatially separate systems. Two systems spatially separate verify the condition about do not have space points in common. Two systems are isolated if between both of there are not exchange energy or matter. Two systems that verify this two qualities can have a common causal origin.

In the example proposed by D. Bohm [16] a singlet state of two fermions can disintegrate in its two constituent particles of half-integral spin with same probabilities each way for have the fermion with the positive or negative spin. When measuring one of the particles, their wave function is "collapses" and automatically most be collapsed the wave function of the other particle, so that, both of they show contrary signs of spin. The instantaneous collapse of both wave functions assures the conservation of the spin magnitude. The paradox in this process is: "how can it have travelled the information of the collapse of the wave functions and its spin value until the other one, even although this can suppose to travel quicker than the light?.

From the causal point of view do not exist paradox. All the systems that we describe in our process has common causal connection in some point.

Let us reconstruct the experiment *ab initio*: In the instant $t = 0$ a scientist enters in the laboratory and he decides that he will carry out the experiment of Bohm to check the speed of the collapse of the wave function. For it, in $t = 1$ hour he has placed the measure instruments (two Stern-Gerlach imams) and in $t = 2$ hours he has prepared the singlet system with two fermions, awaiting that the singlet itself disintegrates. In $t = 2$ hours and 1 minute the singlet itself disintegrate, in $t = 2$ hours, 1 minute and 0,00001 seconds both detectors take the measure of the spin of both particles product of the disintegration and in $t = 2$ hours, 1minute and 10 seconds, the scientist reads the results of both measures and concludes that the collapse of the wave function took place instantly in both particles, without the light could arrive from one to other and coinciding with the measures of the expected results. How can it explain this from the causal point of view?.

The whole exposed process has a common causally point to all the subsystems. This is the instant $t = 0$. From that instant we can isolate each subsystem, we can separate they spatially, but they will not be, in strict sense, independent systems. The description of the preparation of the singlet that disintegrates is not complete if we do not bear in mind the description of the previous steps as they are the placement of the measure instruments or the facts made by the scientist.

A particle, singlet or quantum system, as system, is described by their action (that will conduct to a wave function), but the complete system must be contain the action of all the subsystems that have common causally origin. The action of the system will be the sum of the actions of the particles, the measure instruments, the scientist, and a many more terms besides for each subsystem with common causal origin.

In the fact of the measure it is manifested the final result of the sum of the actions of the different subsystems. When we do not bear in mind the rest of the subsystems that intervene in the process, we are forced to say that, in the measure, the wave function is collapses, but the concept of "collapse" comes to substitute to the: result of the interaction of all the subsystems with common causal origin that are present in the measure process.

Although we argued that the measure instruments cannot have an common causally origin and the two particles of the disintegration can travel in contrary senses one regarding the other one, to the light speed, so that we can obtain the results of both measures and to compare them, both measuring instruments, together with the rest of all the process subsystems will be connected causally in some moment.

In summary, two isolated systems are independent, that is to say, they have not a common causal origin, they are not only spatially separate but SPACE-TIMELY separate. Under these conditions it is verified in the totally the principle of causality.

## Summary

The experience and the principle of causality are two facts which go hand in hand in a narrow form. The experience is which allows us to recognize more or less general principles, but without the principle of causality we could not accumulate experience, because the universe would be we irregular. Two observations: 1ª the existence of the 4 dimensions space-time and 2ª the principle of causality, they allow us to conclude the light speed constancy in the inertial frames, as well as the particular characteristics of the light: particles without mass, neither electrical charging, ... and in general without measurable magnitudes in the homogeneous space-time.

The exact formulation of the principle of causality conduct us to understand the meaning of the time coordinate, its identity with the space coordinates and lastly to solve the paradox that raise the local action principle.